\documentstyle[preprint,revtex]{aps}	% for preprint
\newcommand{\sn}{{\rm sn}}
\newcommand{\sgn}{{\rm sgn}}
\newcommand{\lqm}{{[\!|}}
\newcommand{\rqm}{{|\!]}}
\newcommand{\dint}{{\rm d}}

\begin{document}
\draft

%%%%%%%%%%%%%%%%%%%%%%%%%%%%%%%%%%%%%%%%%%%%%%%%%%%%%%%%%%%%%%%%%%%%%
% title page
%%%%%%%%%%%%%%%%%%%%%%%%%%%%%%%%%%%%%%%%%%%%%%%%%%%%%%%%%%%%%%%%%%%%%
\begin{title}
Solution of Some Integrable One-Dimensional Quantum Systems
\end{title}

\author{Bill Sutherland}
\begin{instit}
Department of Physics\\
University of Utah\\
Salt Lake City, Utah 84112
\end{instit}

\author{B.\ Sriram Shastry}
\begin{instit}
AT\&T Bell Laboratories\\
Murray Hill, New Jersey 07974
\end{instit}

%%%%%%%%%%%%%%%%%%%%%%%%%%%%%%%%%%%%%%%%%%%%%%%%%%%%%%%%%%%%%%%%%%%%%
% add 'Received' later by typesetter
%%%%%%%%%%%%%%%%%%%%%%%%%%%%%%%%%%%%%%%%%%%%%%%%%%%%%%%%%%%%%%%%%%%%%
\receipt{}

%%%%%%%%%%%%%%%%%%%%%%%%%%%%%%%%%%%%%%%%%%%%%%%%%%%%%%%%%%%%%%%%%%%%%
% Abstract
%%%%%%%%%%%%%%%%%%%%%%%%%%%%%%%%%%%%%%%%%%%%%%%%%%%%%%%%%%%%%%%%%%%%%
\begin{abstract}
In this paper, we investigate a family of one-dimensional multi-component
quantum many-body systems.
The interaction is an exchange interaction based on the familiar family of
integrable systems which
includes the inverse square potential.  We show these systems to be integrable,
and exploit this
integrability to completely determine the spectrum including degeneracy, and
thus the
thermodynamics.  The periodic inverse square case is worked out explicitly.
Next, we show that in
the limit of strong interaction the "spin" degrees of freedom decouple.  Taking
this limit for our
example, we obtain a complete solution to a lattice system introduced recently
by Shastry, and
Haldane;  our solution reproduces the numerical results.  Finally, we emphasize
the simple
explanation for the high multiplicities found in this model.
\end{abstract}
\newpage

%%%%%%%%%%%%%%%%%%%%%%%%%%%%%%%%%%%%%%%%%%%%%%%%%%%%%%%%%%%%%%%%%%%%%
% Here goes the main text body
%%%%%%%%%%%%%%%%%%%%%%%%%%%%%%%%%%%%%%%%%%%%%%%%%%%%%%%%%%%%%%%%%%%%%
The plan of this paper is to first review integrable systems and the Lax
technique, show that certain
modified systems with an exchange interaction are integrable, give the general
solution to these
systems and an explicit solution for an important example, then to produce from
this solution by an
appropriate limit the solution to a family of lattice problems, and finally we
discuss what it all
means.  The impatient reader may skip to the closing paragraphs for a summary
of the results.

\section{Integrals of Motion}

If we consider a classical system of N one-dimensional particles governed by
the Hamiltonian
$$H={1
\over 2}\sum\limits_{1\le j\le N}^{} {p_j^2}+\sum\limits_{1\le j<k\le N}^{}
{v(x_j-x_k),}$$ then it
has been shown by Lax  \cite{1}, Moser \cite{2} and Calogero \cite{3} that for
certain potentials
one can find two
Hermitean $N \times N$ matrices $L$ and $A$ that obey the Lax equation
$$\dint L / \dint t=i[A,L].$$
Thus $L$ evolves by a unitary transformation generated by $A$, and hence
 $\det[L - wI]$ is a constant of
motion.  Expanding the determinant in powers of $\omega$ we find $N$ integrals
of motion$$\det [L-\omega
I]=\sum\limits_{0\le j\le N}^{} {J_j(-\omega )^{N-j}},$$ j=1,...,N.  Further,
these integrals have
been shown to be in involution, and thus the system is integrable.

The two matrices are given as
\begin{eqnarray*}
A_{jk} &= &\delta _{jk}\sum\limits_{l(\ne j)} {\gamma
(x_j-x_l)}+(1-\delta _{jk})\beta (x_j-x_k),\\
L_{jk} &= &\delta _{jk}p_j+i(1-\delta _{jk})\alpha (x_j-x_k).
\end{eqnarray*}
The functions v(x), $\gamma$(x) and
$\beta$(x) are even, while $\alpha$(x) is odd.  They obey the equations,
\begin{eqnarray*}
v(x)           &= &-2\alpha(x)\beta (x),\\
\beta(x)       &= &-\alpha (x),\\
\alpha(x+y)[\gamma (y)-\gamma (x)]
               &= &\alpha (y)\alpha (x)-\alpha (x)\alpha (y).
\end{eqnarray*}
Calogero \cite{3}
has shown that the most general solution to these equations is given in terms
of the Jacobi elliptic
function $\sn(x|m)$ as
\begin{eqnarray*}
\alpha (x|m,\lambda ,\kappa )
 &= &\lambda \alpha (\kappa x)={\lambda \over {\sn(\kappa x|m)}},\\
\gamma (x|m,\lambda ,\kappa ,c)
 &= &\lambda \kappa \gamma (\kappa x)+c=\lambda \kappa \alpha ^2(\kappa
x)+c={{\lambda \kappa } \over {\sn^2(\kappa x|m)}}+c,\\
v(x)
 &= &\lambda ^2\alpha ^2(\kappa x)+v_o.
\end{eqnarray*}
Calogero \cite{4} has also demonstrated that if one
replaces the classical dynamical variables with the corresponding quantum
mechanical operators,
$\det[L-wI]$ is well defined with no ordering ambiguity, and the quantum
mechanical commutator  $\lqm H,\det[L-wI] \rqm=0$.  Thus, the $J_j$ are still
constants
of motion.  Finally, Calogero showed that
 $\lqm\det[L-wI], \det[L-w'I] \rqm=0$, and thus the quantum system is also
completely
integrable.

	To sum up the situation:  The one dimensional quantum system, governed by the
Hamiltonian $$H={1
\over 2}\sum\limits_{1\le j\le N}^{} {p_j^2}+\sum\limits_{1\le j<k\le N}^{}
{{{\lambda ^2} \over
{\sn^2(\kappa x|m)}},}$$ is completely integrable.  For the general classical
system integrability
tells us something concrete, namely that the motion in terms of action-angle
variables is on a
torus.  However, for the general quantum system integrability seems to buy one
almost nothing.  The
exception is for those special cases which support scattering; i.e., systems
which fly apart when the
walls of the box are removed.    In these cases, in the distant past and future
the Lax matrix $L$
approaches a diagonal matrix, so that  $$\det [L-\omega I]=\prod\limits_{1\le
j\le N} {(p_j-\omega
)}.$$ Thus the individual momenta $p_{j}$ are conserved in a collision, and
hence
the wavefunction is
given asymptotically by Bethe's ansatz.  Sutherland \cite{5} has exploited this
fact
to completely
determine, in the thermodynamic limit, properties of systems interacting by
potentials of the forms
 $v(x) = g/x^2, g/\sin^2(x)$ and $g/\sinh^2(x)$, including the Toda lattice.
We
emphasize that no features of
the proof of integrability are needed.  All we need is to know it to be
integrable, by whatever
method. (We might even just assume it to be integrable, and deduce the
consequences.)

\section{A Modified Problem}

	Although the previous problems are multi-component problems, in fact
statistics enters only
trivially, due to the strong repulsion of the potential at the origin.
Polychronakos \cite{6} has modified
this problem \cite{10,11}, allowing particles to penetrate by means of an
exchange
interaction.  He then
proves the integrability of this modified problem.  We now offer an alternative
proof of
integrability;  we extend and elaborate on this method in a separate paper
\cite{12}.
 If $P_{jk}$ is the
permutation operator that permutes particles $j$ and $k$, let us modify the Lax
matrices to be
\begin{eqnarray}
A_{jk}
 &= &\delta _{jk}\sum\limits_{l(\ne j)} {P_{jl}\gamma
(x_j-x_l)}+(1-\delta
_{jk})P_{jk}\beta (x_j-x_k),\\
L_{jk}
 &= &\delta _{jk}p_j+i(1-\delta _{jk})P_{jk}\alpha (x_j-x_k).
\end{eqnarray}
We assume that the new potential $v_{jk}$ commutes with $P_{jk}$.  We then seek
to
satisfy a quantum Lax
equation of the form  $\lqm H,L \rqm = [L,A]$.  Here the first fancy commutator
is a
quantum mechanical
commutator between operators, while the second commutator is an ordered matrix
commutator, so that
the equation above is really $N^{2}$ equations of the form
\begin{displaymath}
[H,L_{jk}]=\sum\limits_{1\le l\le N}
{(L_{jl}A_{lk}-A_{jl}L_{lk})}.
\end{displaymath}
  In order that these equations be satisfied, one finds that the functions
 $\alpha(x), \beta(x)$ and $\gamma(x)$ must
satisfy exactly the same equations as before.  The potential now is
\begin{displaymath}
v_{jk}(x)
= \lambda^{2} \alpha^{2}(\kappa x)
 + \lambda\kappa\gamma(\kappa x)P_{jk}
= {{\lambda ^2+\lambda \kappa P_{jk}} \over
{\sn^2(\kappa x | m)}} .
\end{displaymath}

	We will henceforth be interested in a particular case.
Let $\kappa = i, \lambda = i, c = -1$ and $m \rightarrow 1$, so that
\begin{eqnarray*}
\alpha(x)  &\rightarrow &\cot(x),\\
\gamma(x)  &\rightarrow &-1 / \sin^2(x) = -1-\alpha^2(x),\\
\alpha'(x) = -\beta (x)&\rightarrow &-1 / \sin ^2(x)=\gamma (x).
\end{eqnarray*}
Restoring scale factors, we have a potential   $$v_{jk}(x)=(\lambda ^2-\lambda
P_{jk})\kappa ^2/
\sinh ^2(\kappa x).$$
(We could as well have a sine instead of a sinh.)

The significance of this case is for the form of the Lax $A$ matrix.  Since
 $\beta(x) = -\gamma(x)$, then
defining a vector $\eta$ with $\eta_{j}=1$, we see
 $A \eta = \eta^{\dagger} A = 0$.
This allows us to construct constants of motion
by $I_{n} = \eta^\dagger L_{n} \eta$, since
\begin{eqnarray*}
\lqm H,I_n\rqm
 &= &\eta^\dagger \lqm H,L^n \rqm \eta
  =  \eta^\dagger \sum_{0<j<N-1} \{ L^j \lqm H,L \rqm L^{N-1-j} \} \eta \\
 &= &\eta^\dagger \sum_{0<j<N-1} \{ L^j [ A,L ] L^{N-1-j} \} \eta
  =  \eta^\dagger \{ A L^{N-1}-L^{N-1} A \} \eta =0.
\end{eqnarray*}
By Jacobi's relation for commutators,
 $\lqm I_{n},I_{m}\rqm$ is a constant of motion, and
since this is a system
that supports scattering, we see $\lqm I_{n},I_{m} \rqm \rightarrow 0$, and
hence the system is
completely integrable.  Thus
once again, the individual momenta $p_{j}$ are conserved in a collision, and
hence
the wavefunction is
given asymptotically by Bethe's ansatz.  This allows one to completely
determine, in the
thermodynamic limit, properties of systems interacting by potentials
\begin{displaymath}
v(x) =
(\lambda^{2}-\lambda P)/x^{2}
\end{displaymath}
and
\begin{displaymath}
(\lambda^{2}-\lambda P)/\sinh^{2}(x),
\end{displaymath}
as well as the periodic versions of these such as
\begin{displaymath}
(\lambda^{2}-\lambda P) \pi^{2} /L^{2} \sin^{2} (\pi x/L)
\end{displaymath}
in the limit as $L\rightarrow\infty$, density finite.

\section{The \mbox{$(\lambda^2-\lambda P)/x^2$}
and \mbox{$(\lambda^2-\lambda P)\pi^2/L^2\sin^2(\pi x/L)$} Problems}

	The input for the Bethe ansatz wavefunction is the two-body scattering matrix,
since the $N$-body
scattering is simply a product of all $N(N-1)/2$ scatterings.  At the same
time,
we can gain some
confidence by confirming the consistency of the two-body scattering matrix;  it
must satisy the
Yang-Baxter equations.  With the relative coordinate $r = x_{2} - x_{1}$,
the potential
is
\begin{displaymath}
v(r)=(\lambda^{2}-\lambda P)/r^{2},
\end{displaymath}
so depending on whether the
wavefunction is even or odd, the
potential is
$v_\pm(r) = \lambda(\lambda\mp 1)/r^{2}$.
The radial equation then is $-\varphi'' +\sigma
(\sigma -1)\varphi /
r^2=k^2\varphi ,$
  with $\sigma= \lambda$ or $\lambda+1$ according to whether the wavefunction
is even or odd.  The
solution is given as
\begin{displaymath}
\begin{array}{llll}
\varphi (r)
  &=           &\sqrt rJ_{\sigma -1 / 2}(kr), & r>0,\\
  &\rightarrow &\cos (kr-\pi \sigma / 2),     & r\rightarrow \infty \\
  &\rightarrow &r^\sigma,                     & r\rightarrow 0.
\end{array}
\end{displaymath}
For the even and odd solutions then we have  $\varphi _\pm (r)\rightarrow
e^{-ikr}\pm
e^{-i\pi \lambda
}e^{ikr},\quad r\rightarrow \infty .$

To construct a scattering state in which particle 1 with momentum $k_{1}$ is
incident on particle 2  with
momentum $k_2, k_1>k_2$, we take the linear combination
\begin{displaymath}
\begin{array}{llll}
{1 \over {\sqrt 2}}[\psi _+ +\psi_-]
 &\rightarrow &e^{i(k_1x_1+k_2x_2)},                  &x_1<<x_2,\\
 &\rightarrow &e^{-i\pi \lambda}e^{i(k_1x_1+k_2x_2)}, &x_2<<x_1.
\end{array}
\end{displaymath}

Thus, there is no reflection, and the particles pass through each other with
only a phase shift $e^{-i\pi\lambda}$.  Consistency is not an issue.  This is a
surprising result with
important consequences.

	We now place the particles on a large ring of circumference $L$, making the
problem periodic.  We thus
need to replace the potential with some periodic potential, such as
$v(x) = (\lambda^2-\lambda P) \pi^2 / L^2 \sin^2(\pi x/L),$
which reduces to $(\lambda^2-\lambda P)/x^2$ in the limit
$L\rightarrow\infty$, and which therefore has the same
thermodynamic limit.
To repeat: We will be giving the exact solution for the inverse sine squared
exchange potential in
the thermodynamic limit of an infinite system with finite density.

	Taking the $j$th particle around the ring of circumference $L$, it scatters
from
every other particle
with a net phase shift that must be unity, to insure that the wave function is
periodic.  Thus the
$k$'s satisfy the equation $\exp[ik_jL-i\pi \lambda \sum\limits_{l(\ne j)}
{\sgn(k_j-k_l)}]=1,$
  or upon taking the logarithm, $k_j=k_j^o+{{\pi \lambda } \over
L}\sum\limits_{l(\ne j)}
{\sgn(k_j-k_l)}.$

Here $k_j^o=2\pi\cdot \mbox{integer}\cdot L$
 are the the non-interacting momenta with $k_1^o\ge k_2^o\ge ...\ge k_N^o$ .
We easily solve for the
$k$'s obtaining $k_j=k_j^o+\pi \lambda (N+1-2j)/ L.$
 Then the energy is  $$E={1 \over 2}\sum\limits_{1\le j\le N} {k_j^2}={1 \over
2}\sum\limits_{1\le
j\le N} {(k_j^o)^2}+{{\pi \lambda } \over L}\sum\limits_{1\le j\le N}
{k_j^o(N+1-2j)}+\pi ^2\lambda
^2N(N^2-1) / 6L^2,$$
while the momentum is $$P=\sum\limits_{1\le j\le N} {k_j^{}}=\sum\limits_{1\le
j\le N} {k_j^o.}$$

All of this is just a repetition of the old results \cite{5} for the usual
$1/r^2$
potential, which in fact is
a special case of this interaction with a single species of particle.

	Although the energy eigenvalues are exactly same, these levels are
now highly degenerate.  In fact,
the degeneracy are exactly the same as the free particle degeneracy when
$\lambda\rightarrow 0$.
Thus, let us introduce
occupation numbers $\nu_\alpha(k)$, equal to the number of particles of species
$\alpha$ with
non-interacting momenta
$k$.  Thus $\nu_\alpha(k) = 0,1$ for fermions, and $\nu_\alpha(k) = 0,1,2,
\ldots$ for bosons.  Then the
energy $E$ is given by
\begin{eqnarray*}
E/L
&= &{1 \over {4\pi }}\sum\limits_\alpha  {}\int_{-\infty
}^\infty  {\dint k\ \nu
_\alpha (k)k^2}+{\lambda  \over {8\pi }}\sum\limits_{\alpha ,\beta }
{}\int_{-\infty }^\infty  {\dint k\ \nu
_\alpha (k)}\int_{-\infty }^\infty  {\dint k'\ \nu _\beta (k')|k-k'|}+\pi
^2\lambda
^2d^3/ 6\\
    &= &{1 \over {4\pi }}\int_{-\infty }^\infty  {\dint k\ \nu (k)k^2}+{\lambda
\over {8\pi
}}\int_{-\infty }^\infty  {\dint k\ \nu (k)}\int_{-\infty }^\infty  {\dint k'\
\nu(k')|k-k'|}+\pi ^2\lambda ^2d^3/6
\end{eqnarray*}
Here, the density of species $\alpha$ is $d_\alpha = N_\alpha/L = {1 \over
{2\pi }}\int_{-\infty
}^\infty  {\dint k\ \nu
_\alpha (k)}$ , $\nu (k)=\sum\limits_\alpha  {\nu _\alpha (k)}$ , and  $d =
N/L$
= $\sum\limits_\alpha
{d_\alpha }
 = {1 \over {2\pi }}\int_{-\infty }^\infty  {\dint k\ \nu (k)}$
 is the total density.  The degeneracies are now expressed in terms of the
usual entropy $S$, as  $$S/
L=(1/ 2\pi )\sum\limits_\alpha  {}\int_{-\infty }^\infty  {\dint k\ }(\pm
1)(1\pm \nu
_\alpha )\ln (1\pm
\nu _\alpha )-\nu _\alpha \ln (\nu _\alpha ).$$ (Here, and in what follows,
the upper sign is for
bosons, while the lower sign is for fermions.)

We can now determine the equilibrium occupation numbers $\nu_\alpha (k)$ as
those which
maximize
$$
	P/T = S/L - E/TL + \sum\limits_\alpha  {\mu _\alpha N_\alpha / TL} .
$$
   This
gives the familiar
expression $\nu _\alpha =1 / [e^{(\varepsilon -\mu _\alpha ) / T}\mp 1].$
  In this expression,  $$\varepsilon =k^2 / 2+(\lambda  / 2)\int_{-\infty
}^\infty  {\dint k'\ \nu(k')}|k-k'|+\pi ^2\lambda ^2d^2 / 2.$$

Differentiating, we find for the interacting momenta $p$ that appear in the
asymptotic Bethe ansatz,  $p
= \epsilon' = k + (\lambda  / 2)\int_{-\infty }^\infty  {\dint k'\
\nu(k')}\sgn[k-k']$ .
Differentiating once again,
$p'=\varepsilon'' =1+\lambda \nu =p{{dp} \over {d\varepsilon }}={d \over
{d\varepsilon }}\left( {{{p^2}
\over 2}} \right).$
  Thus, integrating, we obtain  $p^2 / 2=\varepsilon +\lambda
T\sum\limits_\alpha  {(\pm 1)\ln [1\mp
e^{(\mu _\alpha -\varepsilon ) / T}]}.$
  We may then go back and write the densities as  $2\pi d_\alpha =\int_{-\infty
}^\infty  {dp{{\nu
_\alpha } \over {1+\lambda \nu }}}.$
  This gives us a complete solution for the thermodynamics.

	For instance, if we consider a system consisting of two species of bosons, say
$+$ and $-$, then taking
$\mu_\pm = \mu \pm h$, we find $e^{-p^2 / 2\lambda T}=[e^{(\varepsilon -\mu ) /
T}+\cosh
(h / T)]^2-\sinh ^2(h /
T),$
  so   $e^{(\varepsilon -\mu ) / T}=[e^{-p^2 / 2\lambda T}+\sinh ^2(h /
T)]^{1/ 2}-\cosh (h / T)$ .
This then allows us to write $\nu_\pm$ as an explicit function of $p$, which
may be
integrated to give $d_\pm$ as a
function of $h_\pm$, and which in turn can be integrated with respect to
$h_\pm$ to give
the free energy, which
can be referenced to the known one-component system at
$h_+ \rightarrow \infty$.

\section{Ground State Energy}

	At this point, let us look at the ground state energy as a function of the
densities $\{d_\alpha\}$.  This is
easily done, and it is only for the ground state when $T=0$ that we expect
singularities.  We assume $f$
species of fermions with densities
$d_1\leq d_2\leq\ldots\leq d_f$,  and a total of $d_B$ bosons of
any species, so the
total density is $d$.  Then the occupation numbers are the free particle ground
state occupation
numbers, and the integrals in the expression for the energy are easily
evaluated to give $$6E / \pi
^2L=(1+2\lambda )\sum\limits_{1\le \alpha \le f} {d_\alpha ^3}+3\lambda
d_B\sum\limits_{1\le \alpha
\le f} {d_\alpha ^2}+\lambda \sum\limits_{1\le \alpha <\beta \le f} {d_\alpha
(d_\alpha ^2+3d_\beta
^2)+\lambda ^2d^3.}$$

Because this expression is only valid for a particular ordering of the
densities, the ground state
does have singularities whenever $d_\alpha=d_\beta$.  A Taylor expansion gives
the
singularity as $E / L\approx
|d_\beta -d_\alpha |^3,$
 so the susceptibilities do not diverge.  The singularity structure, and even
the nature of the
singularities, is very similar that found previously for a short-ranged lattice
exchange model \cite{7}.

\section{The Lattice Problem}

	Examining the potential energy as $\lambda\rightarrow\infty$, we see that the
particles crystallize
into a lattice with
lattice constant $1/d$, and so the elastic modes of this lattice, and the
compositional or "spin" modes
on the lattice separate.  We thus have as
$\lambda\rightarrow\infty$,
\begin{eqnarray*}
H &\rightarrow &(1 /
2)\sum\limits_{1\le j\le N}
{p_j^2}+\lambda ^2\sum\limits_{1\le j<k\le N} {{{\pi ^2} \over {L^2\sin ^2[\pi
(x_j-x_k) /
L]}}}-\lambda d^2\sum\limits_{1\le j<k\le N} {{{\pi ^2P_{jk}} \over {N^2\sin
^2[\pi (j-k) / N]}}}\\
  &= &H_e+\lambda d^2\,H_s.
\end{eqnarray*}
The Hamiltonian for the elastic modes is just the familiar one-component
system \cite{5}, although with a
coupling constant $\lambda^2$.  We have defined the spin Hamiltonian
$H_s$ with exchange
constant unity, although
the effective exchange constant of the last term of $H$ is
$\lambda d^2$.  The two-species
problem was first
studied independently and simultaneously by Shastry \cite{8}, and by Haldane
\cite{9}.
Thus we see that by an
appropriate limiting process we may determine the thermodynamics of this spin
Hamiltonian exactly.
For the spin problem, the appropriate concentrations are
$m_\alpha = N_\alpha/N = d_\alpha/d.$

	Let $F(T, \{N_\alpha\}, L)$ be the free energy for the original continuum
exchange
problem.  Then it is seen
on dimensional grounds to be of the form
$L T^{3/2} f(\{d_\alpha\}/T^{1/2})$.  Given a set of
"chemical potentials"
${h_\alpha}$, such that  $\sum\limits_\alpha  {h_\alpha }=0$ , then we make a
partial
Legendre transformation,
$$F(T,\{N_\alpha \},L)-\sum\limits_\alpha  {h_\alpha N_\alpha
}=F(T,\{h_\alpha
\},N,L)=LT^{3 / 2}f(d / T^{1 / 2},\{h_\alpha \} / T^{1 / 2}).$$
(Free energies are identified by their variables.)  Then $N_\alpha =-\partial
F / \partial h_\alpha ,$
  so if   $d_\alpha(d,\{h_\alpha\}) = -\partial f(d,\{h_\alpha \})/ \partial
h_\alpha $ , we
can write
$d_\alpha=T^{1/2}d_\alpha(d/T^{1/2}, \{h_\alpha\}/T)$, which for $T=1$ is
simply
$d_\alpha$.

	Returning now to the limit as $\lambda\rightarrow\infty$, and setting
$T = 1$, we have
$f(\{d_\alpha\})\rightarrow f_1 (d) + d f_s (1/\lambda d^2,\{m_\alpha\})$.
Here $f_1(d)$ is the free energy density for the one-component system
\cite{5}
with coupling constant
$\lambda^2$, and $f_s(T, \{m_\alpha\})$ is the free energy per site for the
spin system.  Again
making the partial
Legendre transformation, we find
$$f_s(T,\{h_\alpha\})=\mathop {\lim}\limits_{\lambda \rightarrow \infty}
\sqrt {\lambda T}\left[ {f(1 / \sqrt {\lambda T},\{h_\alpha\} / T)\,-f_1(1 /
\sqrt {\lambda T})\,}
\right].$$

Remember, in the functions on the right-hand side, the first arguement
represents the density
$d =
1 / \sqrt {\lambda T}
 \rightarrow 0$ as $\lambda\rightarrow\infty$.  Finally, the relationship is a
little clearer in terms of the
concentrations,
$m_\alpha =\mathop {\lim}\limits_{\lambda \rightarrow \infty }
d_\alpha / d(\{\mu_\alpha \})$ , where the
chemical potentials are $\{\mu_\alpha\}=\{\mu + h_\alpha/T\}$, and $\mu$ is
adjusted to give $1=T\lambda d^2$.

	Referring to our previous explicit results, in the limit as
$\lambda\rightarrow\infty$, with
$1=T\lambda d^2$, then  $m_\alpha
=d_\alpha  / d = 1 / 2\pi =\int_{-\pi }^\pi  {dp\,\nu _\alpha  / \nu },$
  with  $\nu _\alpha =1 / \left[ {\exp (\varepsilon -h_\alpha  / T)\mp 1}
\right],$
and  $(p^2-\pi ^2) / 2=T\sum\limits_\alpha  {(\pm 1)\ln [1\mp \exp (h_\alpha  /
T-\varepsilon )]}.$
  This gives us a complete solution, since the concentrations can be integrated
with respect to $\{h_\alpha\}$
to give the free energy.  The free energy is referenced with respect to the
trivial case of $m_\alpha=1$ for
some $\alpha$.  For a two species problem of either both fermions or both
bosons, our
results reproduce
Haldane's conjecture based on his numerical investigation \cite{9}.  We
emphasize
that one changes the sign
of the of the spin Hamiltonian from ferromagnetic to antiferromagnetic simply
by changing the
statistics of the particles.

\section{Summary and Conclusion}

	Let us now summarize what we have accomplished.  We have investigated a family
of one-dimensional
multi-component quantum many-body systems, interacting by an exchange
interaction based on the
familiar family of integrable systems which includes the inverse square
potential.  These systems are
shown to be integrable.  Since these systems support scattering, the asymptotic
wave function is
given by Bethe's ansatz, and consequently the spectrum can be completely
determined, in the
thermodynamic limit.  The interactions that can be solved in this way include
$v(x) = (\lambda^2-\lambda P)/x^2$
and
$(\lambda^2-\lambda P)/\sinh^2(x)$ at finite density, as well as the periodic
versions of these such as
$(\lambda^2-\lambda P) \pi^2/L^2\sin^2(\pi x/L)$ in the limit as
$L\rightarrow\infty$, density finite.

	We explicitly work out the case of the inverse square or inverse sine square
potentials, and
demonstrate that for this exchange potential particles scatter without
reflection.  They pass through
one another with only a phase shift, and this phase shift is independent of the
identities of the
particles (and the momentum as well).  As a consequence of this perfect
transmission, there can be no
Bragg reflection to split the energy of the even and odd solutions when the
particles pass around the
ring.  Instead we have degenerate states, and so we could take a linear
combination which, for a
two-component system, would have all particles of one type going clockwise
around the ring, while all
particles of the other type go counter-clockwise.  Yet this state would be an
eigenstate of energy.
This then explains the very high degeneracy observed in the spectrum; it is
just the degeneracy of
the free particle system.  We can offer other examples of systems without
reflection, with similar
behavior.  In fact, it is clear from the discussion, that by using adiabatic
continuity, the quantum
numbers of the states can be taken as the corresponding quantum numbers of the
free particle states.
Singularities occur only in the ground state, and these are exhibited.
	Finally, it is demonstrated that for each of these systems, taking the limit
of infinite coupling
constant appropriately gives the solution to a corresponding lattice problem.
One could easily prove
the integrability of these lattice systems directly using the Lax technique.
However, application of
the asymptotic Bethe ansatz directly to the lattice is plagued by umklapp
problems; the momenta are
pushed by the interaction outside the Brillouin zone \cite{10}.  The limiting
trick
avoids these
complications.  Taking this limit in our explicit example, we obtain a complete
solution to a
corresponding lattice system introduced recently by Haldane, and by Shastry.
Our solution reproduces
Haldane's numerical results.  This system exhibits very high multiplicities;
we find these
multiplicities to be just the familiar degeneracy of free particles.  It arises
naturally from the
absence of reflection, and hence of Bragg reflection and the consequent
splitting of levels, and is
nothing more.  This seems to us a good explanation.

	The success of the asymptotic Bethe ansatz has been rather dramatic, so let us
here reiterate the
philosophy behind the method.  To justify using the asymptotic Bethe ansatz,
one needs all of the
following:  a system which supports scattering, a proof of integrability with
integrals that depend
only on the asymptotic momenta, and a valid virial expansion.  Given these, and
using only the
two-body scattering data as input, one can deduce all thermodynamic properties,
including ground
state properties and low-lying excitations, within the phase for which the
virial expansion is
valid.  Often other phases may be reached by symmetry.

	We conclude by reemphasizing that there are many other interesting systems to
be explored in detail
in the future.

\end{document}